# Optimization of compact Soller collimator

L.I.Ognev

RRC «Kurchatov Institute»

## Abstract

The opportunity of optimization compact collimator for soft x-ray radiation for wave length 4 nm, corresponding to carbon atoms absorption was investigated at the specified angle of collimation of the beam. As an example of materials for collimator glass and nickel have been chosen. Both absorption at reflection and scattering on roughness were taken into consideration.

From the obtained results follows that effect of diffraction becomes significant for minimizing width of the channel up to 20 - 40 µm. Reduction of transmission angle from 0.014 up to 0.007 radian results in increase of factor of attenuation of intensity of the beam $2(\beta_{incoh}+\beta_{absorp})L$ for the basic waveguide mode approximately by 5 times. The choice of lighter material for collimator allows to decrease considerably losses at transmission of the basic mode, thus the replacement of nickel with glass decreases losses approximately by 4 times. So in short collimator where angular collimation is achieved on 2 reflections at greater height of roughness of a surface losses are 20% lower, than for more smooth surfaces with 4 beam reflections.

During heating hydrogen plasma in tokamak because of degradation of a graphite covering of a wall dust carbon particles probably penetrate into active zone. Research on peripheral part of the active zone with the method of Rayleigh scattering of laser radiation [1] has shown, that the sizes of particles are in the range of 80 nanometers. However research of the center of the active zone is complicated by absorption of laser radiation in plasma. The technique of research by means of soft x-ray radiation in a range of absorption of atoms of carbon 280 eV [2] is supposed to be more selective. However the geometry of thermonuclear installation makes strict demands to possible dimensions of X-ray optics devices. In the present work the opportunity of miniaturization of Soller collimators for formation and research of x-ray beams in the field of 280 eV was investigated.

Schematically the ray paths in flat collimator are shown in a Fig. 1.

The angular spectrum of the outgoing beam is defined by losses at reflection of beams from collimator walls. From figure it is obvious, that if the ratio between width of the channel *d* and its length *L* remains constant *d / L = const* number of reflections in the geometrical optics approximation is the same and thus losses do not depend on the collimator length. We shall consider, that the factor of reflection can be defined by the formula

$$R(\vartheta) = \exp(-(4\pi\sigma\vartheta/\lambda)^2) \cdot R_F(\vartheta) \qquad (1)$$

Where σ is root-mean-square height of roughness of collimator walls,
λ is a wave length of radiation,

$R_F(\theta)$ is coefficient of the total external reflection of radiation from a surface for a beam with sliding incidence angle θ (Fresnel formula).

The typical dependence $R_F(\theta)$ for reflection of radiation with wave length of λ = 4 nm from glass and nickel surfaces with various values of root-mean-square amplitude of roughness is shown on Fig.2 and 3. Values of optical constants of materials were taken from [2].

For restriction of a beam angular halfwidth $\theta_{1/2}$ on one reflection it is necessary, that $R(\theta_{1/2}) \leq 1/e$. If 2 reflections from walls supposed, the condition should be $\{R(\theta_{1/2})\}^2 \leq 1/e$, for 3 reflections condition $\{R(\theta_{1/2})\}^3 \leq 1/e$ should be satisfied. For $n$ reflections

$$\{R(\theta_{1/2,n})\}^n \leq 1/e. \tag{2}$$

That is, the less is the coefficient of reflection the smaller number of reflections $n$ is required for restriction of angular spectrum of a beam. Thus values of the ratio of the width to length $d/L$ can be obtained.

Height of roughness of a surface of the channel, necessary to satisfy the conditions (2) with $n$ reflections at $\theta_{1/2,n} < \theta_F$, where $\theta_F = \sqrt{(1-\mathrm{Re}(\varepsilon))}$ is Fresnel angle of the total external reflection, can be obtained from a formula

$$\sigma = \frac{\lambda}{4\pi\theta_{1/2,n}}\sqrt{\ln(R_F(\theta_{1/2,n})) + \frac{1}{n}}. \tag{3}$$

For narrow beams the diffraction effects can be expected which results in attenuation of the radiation propagating in the channel collimator even under a zero angle to an axis [3, 4]. The attenuation coefficient of the lower mode due to incoherent scattering on the rough surface at small values of factor of correlation R (z) is proportional to root-mean-square height of roughness [3]

$$\beta_{incoh} \sim k^2(1-\varepsilon)^2\sigma \int_{-\infty}^{\infty} dz' \int_{-\infty}^{\infty} \exp(-\xi^2/2)d\xi \int_{0}^{V(z',\xi)} \exp(-\eta^2/2)d\eta, \qquad (4)$$

where $V(z',\xi) = -R(z')\xi/(1-R^2(z'))^{1/2}$.

Using the approximated approach for wave eigenfunctions, the coefficient of incoherent attenuation can be written approximately as

$$(\beta_{incoh})_l \approx \frac{1}{2}\frac{\pi^2}{d^3}(l+1)^2 \sigma \sqrt{\pi}(\text{Re}(\varepsilon)) \cdot I, \qquad (5)$$

where $l$ is number of waveguide eigenfunction,
$I$ is correlation functional in (2), dependent on correlation properties of a surface of the channel. For exponential autocorrelation function $R(z) = \exp(-z/l_{corr})$ the values of this correlation functional resulted in Table 1.

Table 1

| $l_{corr}$ [µm] | $I/2$ [µm] |
|---|---|
| 0.1 | $0.7086 \cdot 10^{-2}$ |
| 0.5 | $0.3463 \cdot 10^{-1}$ |
| 1.0 | $0.6918 \cdot 10^{-1}$ |
| 5.0 | 0.3399 |
| 10 | 0.6053 |

Dependence of factor of attenuation of a beam on of the channel width due to scattering on rough walls resulted in Fig. 4 for glass and Fig. 5 for nickel at various heights of wall roughness. The length of correlation has been taken equal to 0.1 µm.

Absorption on walls leads to attenuation of radiation for waveguide mode number $l$ with coefficient

$$(\beta_{absorp})_l \approx \frac{\pi^2}{d^3}(l+1)^2 \operatorname{Im}(\varepsilon)[\operatorname{Re}(1-\varepsilon)]^{-3/2}. \tag{6}$$

The approximate formulas (5) and (6) agreed well with results of numerical calculations [3] and explicitly show reverse cubic dependence of factors of attenuation of waveguide modes on width of the channel, earlier noted in numerical calculations [4]. It can be noted also that factor of absorption (4) has the same dependence on parameters of environment as linear approximation of reflection coefficient [5]. Dependence of factor of attenuation of a beam with the wave length λ = 4 nm on width of the channel due to absorption on collimator walls, calculated with these formulas is shown on Fig. 6 for glass and on Fig. 7 for nickel.

Let us consider consequences of miniaturization for losses of radiation at passage through collimator at axial orientation of a beam. We shall consider dependence of losses $2(\beta_{incoh} + \beta_{absorp})L$ on width of the channel $d$ under condition of the constant ratio $d/L$. That corresponds to constant angular transmission spectrum in geometrical optics approach. It results in approximate equation $L = d\,n\,/\,\theta_{½,\,n}$.

The maximum angular range $\theta_{½,n}$ for detection of dust was determined from the expected size of particles of carbon D=60 nm, that is $\theta_½$ ~0.1 λ /D ~7·10$^{-3}$ or 14·10$^{-3}$ rad.

From ratio (2) we determine height of roughness of collimator surfaces which provides attenuation of X-ray radiation with wavelength 4 nm after n=2 and 4 reflections in case of usage of nickel or glass. Results are given in Table 2.

Table 2

|          | Nickel | n=2        | n=4   |
|----------|--------|------------|-------|
| angle of collimation 0.007 | | σ =310.7 | 211.8 |
| 0.014    |        | σ =149.7   | 97.4  |
|          | Glass  | n=2        | n=4   |
| angle of collimation 0.007 | | σ =308.1 | 207.9 |
| 0.014    |        | σ =146.9   | 93.0  |

Despite of triple difference of the value Re(1 - ε) for glass and nickel, amplitudes of height of the roughness necessary for collimation of the radiation, practically coincide. It is accounted for strong influence of roughness on losses of a mirror component of the reflected beam.

These data were used for calculation of losses for the lower wave modes at various values $d/l$. Results of calculations are shown on Fig. 8 and Fig. 9. From the results for wave length 4 nm follows, that effects of diffraction become significant at reduction of width of the channel up to 20 - 40 μm. Reduction of collimation angle from 0.014 up to 0.007 radian leads to increase of attenuation factor $2(\beta_{incoh} + \beta_{absorp})L$ for the basic waveguide mode approximately by 5 times. The choice of lighter material for collimator allows to decrease considerably minimum losses for transmission of radiation through collimator, so replacement of nickel with glass results approximately by 4 times decrease. Thus in short collimator where angular restriction is achieved on 2 reflections at greater height of surface roughness losses are 20 % smaller.

## The main results

1. The approximate analytical expressions for attenuation of wave modes of the X-ray radiation captured in low absorbing dielectric gap with rough walls were developed.

2. Diffraction effects for transmission of X-ray radiation with wavelength of 4 nm become significant when narrowing the width of a channel till 20-40 μm.

3. Choice of a material with smaller nuclear number and the increased roughness of a surface can reduce the minimum losses of the radiation at transmission through the channel at the same angular discrimination of a beam.

# References

1. W.P. West, B.D. Bray, J. Burkart, Measurement of number density and size distribution of dust in DIII-D during normal plasma operation, Plasma Phys. Control. Fusion, 2006, v.48, p. 1661 - 1672.

2. I. Diel, J. Friedrich, C. Kunz, S. Di Fonzo, B. R. Müller, W. Jark, Optical constants of float glass, nickel, and carbon from soft X-ray reflectivity measurements, Applied Optics, 1997, Vol. 36, No. 25, p. 6376 - 6382.

3. T.A. Bobrova, L.I. Ognev, On the "supercollimation" of x-ray beams in rough interfacial channels, JETP Letters, 1999, v.69, n.10, p. 734-738.

4. L.I. Ognev, X-ray diffraction effects in submicron slits and channels, X-ray spectrometry, 2002, vol. 31, n. 3, p. 274-277.

5. A.V. Vinogradov, V.F. Kovalev, I.V. Kozhevnikov, V.V. Pustovalov,
Zh. Tech. Fiz., 1985, v. 55, p. 244;  Sov. Phys. Tech. Phys., 1985, v. 30, 145.

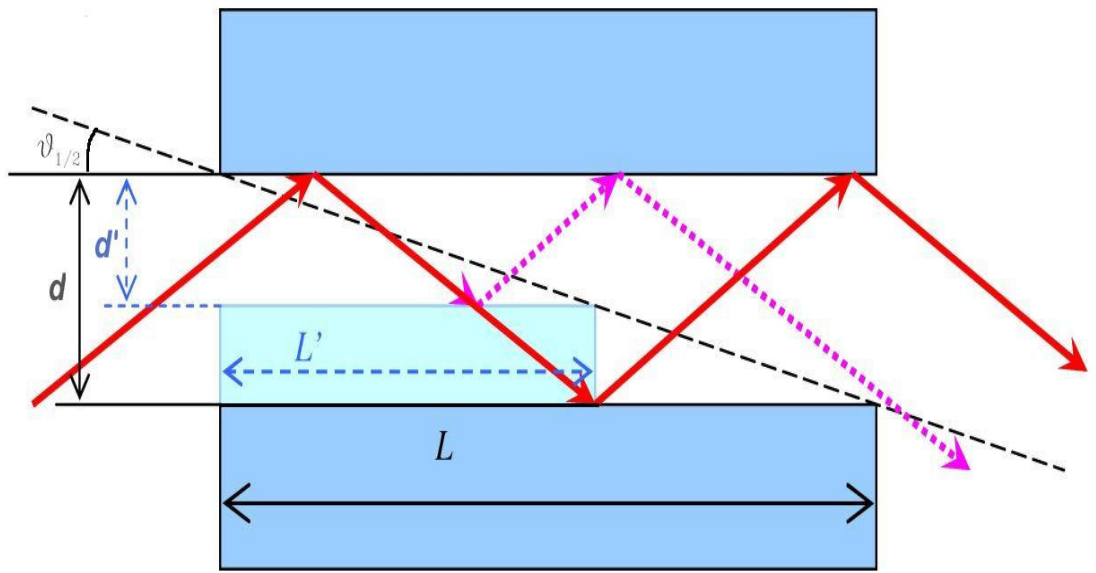

Fig. 1

Changes of the linear size of collimator with given ratio of width $d$ and length $L$ does not change angular spectrum of a target beam in the approximation of geometrical optics.

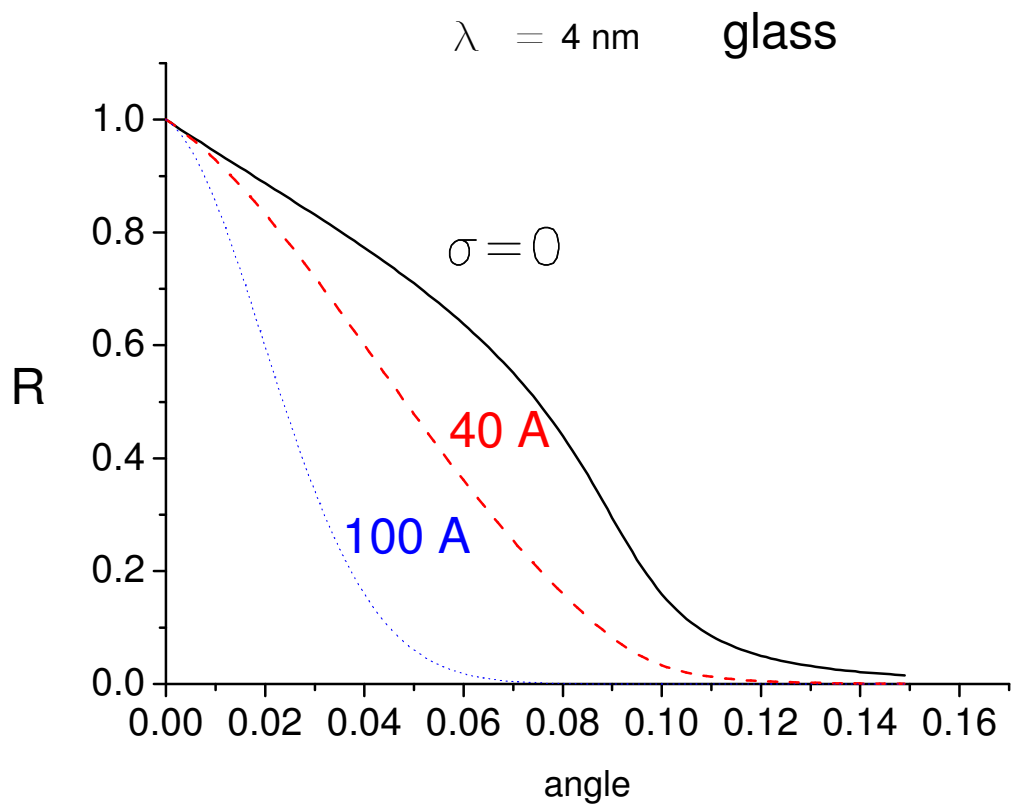

Fig. 2

Dependence of reflection coefficient $R(\theta)$ on sliding angle of incidence θ of radiation with wavelength of 4 nm on glass surface with various values of height of roughness σ = 0 (solid), 40 Å (dashed) and 100 Å (points). For calculation of Fresnel reflection optical parameters of substance were used from work [2].

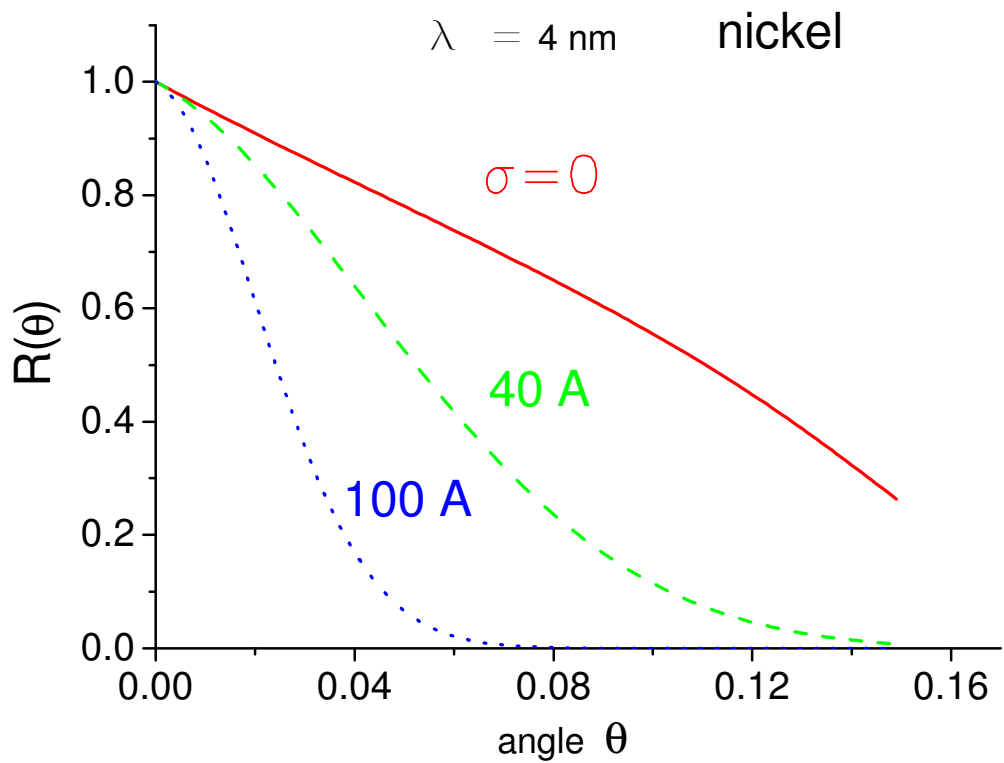

Fig. 3

Dependence of reflection coefficient $R(\theta)$ from a sliding angle of incidence $\theta$ of radiation with wavelength of 4 nm on nickel surface with various values of height of roughness $\sigma$ = 0 (solid), 40 Å (dashed) and 100 Å (points). For calculation of Fresnel reflection optical parameters of substance were used from work [2].

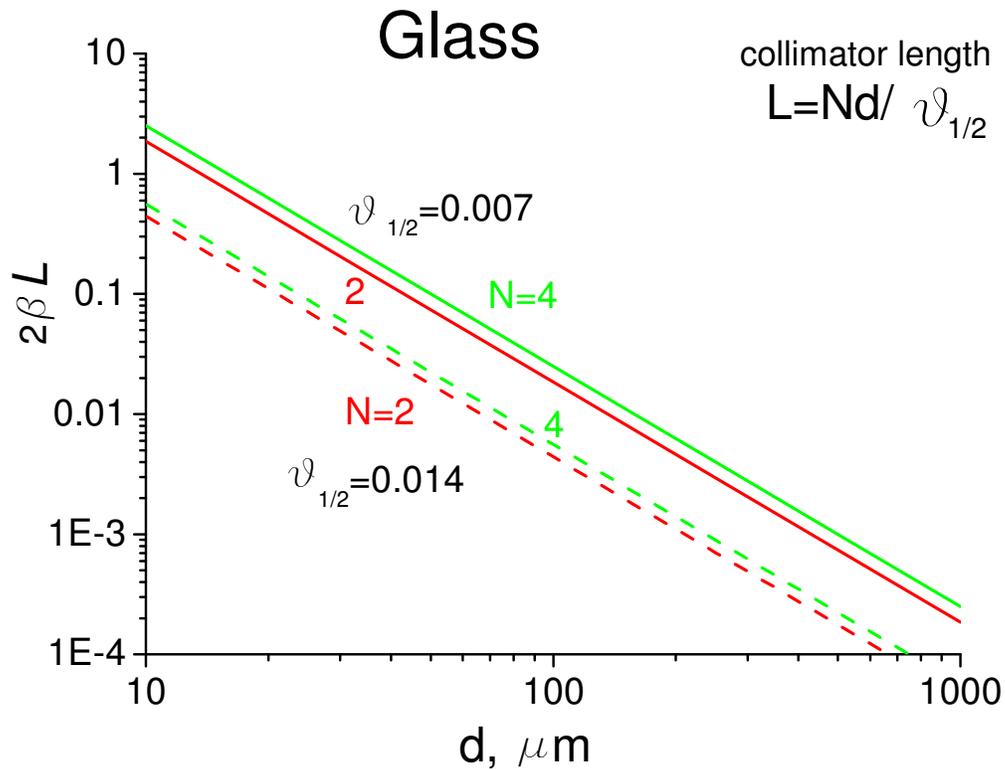

Fig. 4

Dependence of attenuation coefficient of the basic waveguide mode on width of the flat glass channel *d*. The solid line corresponds to attenuation due to absorption in the smooth channel. The dot curve corresponds to attenuation due to scattering on roughness with height σ = 80 Å, dashed line corresponds to σ = 160 Å.

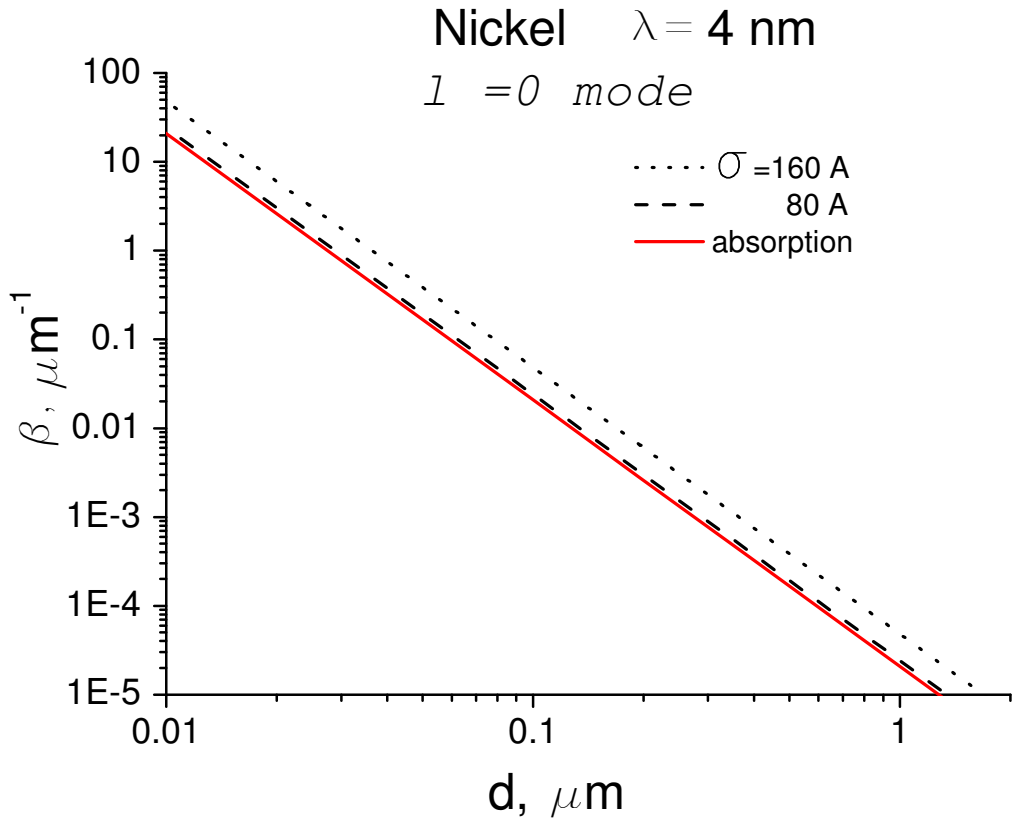

Fig. 5

Dependence of attenuation coefficient of the basic waveguide mode on width of the flat nickel channel *d*. The solid line corresponds to attenuation due to absorption in the smooth channel. The dot curve corresponds to attenuation due to scattering on roughness with height σ = 80 Å, dashed line corresponds to σ = 160 Å.

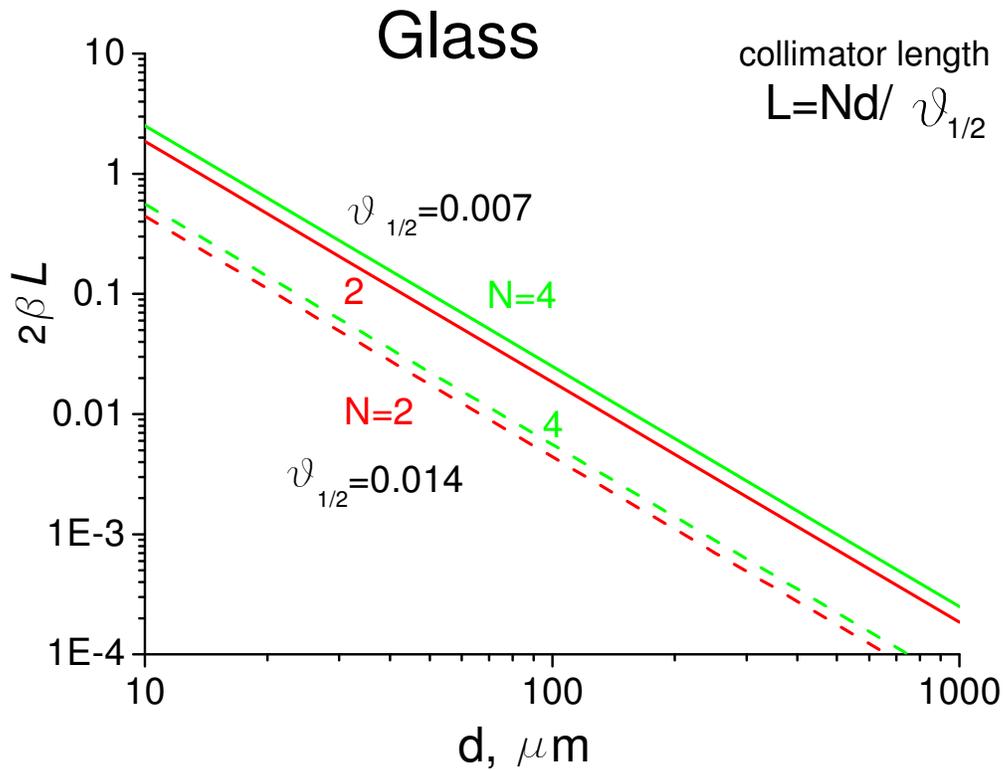

Fig. 6

Dependence of a parameter of attenuation of intensity of radiation $2(\beta_{incoh} + \beta_{absorp})L$ for the basic waveguide mode $l = 0$ on width of a flat glass wave guide $d$. Ratio between collimator length and its width are chosen equal $L/d = N/\theta_{1/2}$, where $\theta_{1/2}$ is half-width collimation angle of a X-ray beam, $N$ is maximum number of reflections. Solid lines above correspond to angular restriction to $\theta_{1/2} = 0.007$, dashed lines to $\theta_{1/2} = 0.014$ at double and quadruple reflections.

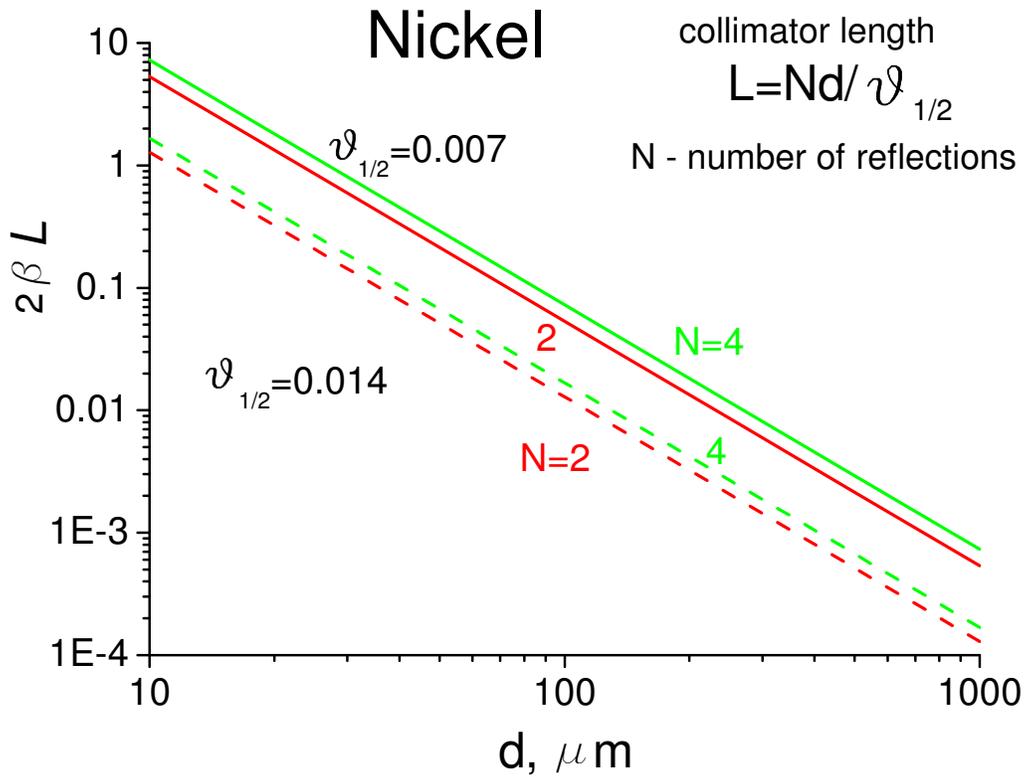

Fig. 7

Dependence of parameter of attenuation of intensity of radiation $2(\beta_{incoh} + \beta_{absorp})L$ for the basic waveguide mode $l = 0$ on width of a flat nickel wave guide $d$. Ratio between collimator length and its width are chosen equal $L/d = N/\theta_{1/2}$, where $\theta_{1/2}$ is halfwidth collimation angle of a X-ray beam, $N$ is maximum number of reflections. Solid lines above correspond to angular restriction to $\theta_{1/2} = 0.007$, dashed lines to $\theta_{1/2} = 0.014$ at double and quadruple reflections.